\documentclass[12pt,preprint]{aastex}
\usepackage{graphicx}
\usepackage{natbib}
\shorttitle{Nuclear constraints on the momenta of inertia of neutron
stars} \shortauthors{Worley et al.}

\begin{document}
\title{Nuclear constraints on the momenta of inertia of neutron stars}
\author{Aaron Worley, Plamen G. Krastev and Bao-An Li} \affil{Department of Physics,
Texas A\&M University-Commerce, Commerce, TX 75429, U.S.A.}
\email{aworley@leo.tamu-commerce.edu,
Plamen\_Krastev@tamu-commerce.edu, Bao-An\_Li@tamu-commerce.edu}

\begin{abstract}
Properties and structure of neutron stars are determined by the
equation of state (EOS) of neutron-rich stellar matter. While the
collective flow and particle production in relativistic heavy-ion
collisions have constrained tightly the EOS of symmetric nuclear
matter up to about five times the normal nuclear matter density,
the more recent experimental data on isospin-diffusion and
isoscaling in heavy-ion collisions at intermediate energies have
constrained considerably the density dependence of the nuclear
symmetry energy at subsaturation densities. Although there are
still many uncertainties and challenges to pin down completely the
EOS of neutron-rich nuclear matter, the heavy-ion reaction
experiments in terrestrial laboratories have limited the EOS of
neutron-rich nuclear matter in a range much narrower than that
spanned by various EOSs currently used in astrophysical studies in
the literature. These nuclear physics constraints could thus
provide more reliable information about properties of neutron
stars. Within well established formalisms using the nuclear
constrained EOSs we study the momenta of inertia of neutron stars.
We put the special emphasis on the component A of the extremely
relativistic double neutron star system PSR J0737-3039. Its moment
of inertia is found to be between $1.30$ and $1.63$
$(\times10^{45}g$ $cm^2)$. Moreover, the transition density at the
crust-core boundary is shown to be in the narrow range of
$\rho_t=[0.091-0.093](fm^{-3})$.
\end{abstract}

\maketitle

\section{Introduction}

Neutron stars exhibit a large array of extreme characteristics.
Their properties and structure are determined by the equation of
state (EOS) of neutron-rich stellar matter at densities up to an
order of magnitude higher than those found in ordinary nuclei, see,
e.g.~\citet{Weber:1999a} and \citet{Lattimer:2004pg}. Therefore, the
detailed knowledge about the EOS of neutron-rich nuclear matter over
a wide range of densities is necessary for the study of neutron
stars. For isospin asymmetric nuclear matter, various theoretical
studies have shown that the energy per nucleon can be well
approximated by
\begin{eqnarray}
E(\rho ,\delta )=E(\rho ,\delta =0)+E_{\mathrm{sym}}(\rho )\delta
^{2}+O(\delta ^{4}),
\end{eqnarray}
in terms of the baryon density $\rho =\rho _{n}+\rho _{p}$, the
isospin asymmetry $\delta=(\rho _{n}-\rho _{p})/(\rho _{n}+\rho
_{p})$, the energy per nucleon in symmetric nuclear matter $ E(\rho
,\delta =0)$, and the bulk nuclear symmetry energy
$E_{\mathrm{sym}}(\rho )$. Here we report the results of a study on
the moment of inertia and the core-crust transition density of
neutron stars within well established formalisms in the literature
using several EOSs constrained by the latest terrestrial heavy-ion
reaction experiments.

Presently, besides the possibilities of phase transitions into
various non-nucleonic states the behavior of nuclear matter under
extreme densities, pressures and/or isospin-asymmetry is still
highly uncertain and relies upon, often, rather controversial
theoretical predictions. This circumstance introduces
corresponding uncertainties in the EOS of neutron-rich nuclear
matter and thus limits our ability to understand many key issues
in astrophysics. While astrophysical observations can also limit
the EOS of neutron-rich nuclear matter, terrestrial laboratories
experiments provide complementary information and have their
unique advantages. In this regard, it is especially interesting to
mention that the collective flow and particle production in
relativistic heavy-ion collisions have constrained the EOS of
symmetric nuclear matter $E(\rho ,\delta =0)$ up to about five
times the normal nuclear matter density to a narrow
range~\citep{Danielewicz:2002pu}. However, there are still many
challenges and uncertainties in pinning down precisely the EOS of
neutron-rich nuclear matter. One of the major remaining
uncertainties is the density dependence of the nuclear symmetry
energy $E_{sym}(\rho)$, see e.g. Refs.
~\citep{Lattimer:2004pg,Steiner:2004fi,Chen:2007} for recent
reviews. To constrain the density dependence of the symmetry
energy, many terrestrial nuclear experiments have been carried out
recently or planned. Depending on the techniques used, some
experiments are more useful for exploring the symmetry energy at
low densities while others are more effective at high densities.
For instance, heavy-ion reactions, especially those involving
radioactive beams, provide a unique means to probe the
$E_{sym}(\rho)$ over a broad density range
~\citep{Li:1997px,Li:2001a,Baran:2004ih,Chen:2007}. In fact, some
significant progress has been made very recently by studying the
isospin
diffusion~\citep{Shi:2003np,Tsang:2004,Chen:2005a,Li:2005jy} and
isoscaling~\citep{Tsang:2001,Shetty:2007} in heavy-ion reactions
at intermediate energies. The analysis of these phenomena based on
transport theories of heavy-ion reactions and thermodynamical
models of nuclear multifragmentation has limited the
$E_{sym}(\rho)$ in a range much narrower than that spanned by
various forms of the $E_{sym}(\rho)$ currently used in
astrophysical studies in the literature. Moreover, the lower bound
of the $E_{sym}(\rho)$ extracted from the heavy-ion reactions is
consistent with the RMF prediction using the FSUGold interaction
that can reproduce not only saturation properties of nuclear
matter but also structure properties and giant resonances of many
finite nuclei~\citep{Piek07}.

It is also well known that the sizes of neutron skins in heavy
nuclei are sensitive to the symmetry energy at subsaturation
densities, see, e.g.,
Refs.~\citep{Brown00,Horowitz:2000xj,Horowitz:2002mb,Die03,Fur02,Steiner:2004fi,ToddRutel:2005fa,Steiner:2005rd,Chen05b}.
However, available data of neutron-skin thickness obtained using
hadronic probes are not accurate enough to constrain significantly
the symmetry energy. Interestingly, the parity radius experiment
(PREX) at the Jefferson Laboratory aiming to measure the neutron
radius in $^{208}Pb$ via parity violating electron scattering
(Jefferson Laboratory Experiment E-00-003) \citep{Horowitz:2001}
hopefully will provide much more precise data and thus constrain the
symmetry energy at low densities more tightly in the near future. On
the other hand, at supranormal densities, a number of potential
probes of the symmetry energy have been
proposed~\citep{Li:1997rc,Li:2000bj,Li:2002qx,Li:1997px,Chen:2007}.
Moreover, several experiments to probe the high density behavior of
the symmetry energy with high energy radioactive beams have been
planned at the CSR/Lanzhou, FAIR/GSI, RIKEN and the NSCL/MSU.

While the EOS of neutron-rich nuclear matter has not been
completely determined yet, it is still very interesting to examine
astrophysical implications of the EOS constrained by the latest
terrestrial laboratory experiments mentioned above. Global
properties of spherically symmetric static (non-rotating) neutron
stars have been studied extensively over many years, for recent
reviews, see, e.g.,
Refs.~\citep{Lattimer:2000kb,Lattimer:2004pg,Prakash:2001rx,Yakovlev:2004iq,Heiselberg:2000dn,Heiselberg:1999mq,Steiner:2004fi}.
However, properties of (rapidly) rotating neutron stars have been
investigated to lesser extent. Models of (rapidly) rotating
neutron stars have been constructed only by several research
groups with various degree of approximation
\citep{Hartle:1967he,Hartle:1968si,Friedman:1986tx,Bombaci:2000rc,1990ApJ...355..241L,1989MNRAS.237..355K,1994ApJ...424..823C,Stergioulas:1994ea,Stergioulas:1997ja,1993A&A...278..421B,1998PhRvD..58j4020B,Weber:1999a,2002A&A...381L..49A}
(see \citet{Stergioulas:2003yp} for a review). In a recent
work~\citep{KLW2} we have reported predictions on the
gravitational masses, radii, maximal rotational (Kepler)
frequencies, and thermal properties of (rapidly) rotating neutron
stars. In this work, using the nuclear constrained EOSs we
calculate the momenta of inertia for both spherically-symmetric
(static) and (rapidly) rotating neutron stars using well
established formalisms in the literature. Such studies are
important and timely as they are related to the astrophysical
observations in the near future. In particular, the moment of
inertia of pulsar $A$ in the extremely relativistic neutron star
binary PSR J0737-3039 \citep{Burgay2003} may be determined in a
few years through detailed measurements of the periastron advance
\citep{BBH2005}.

\section{The equation of state of neutron-rich nuclear matter
constrained by recent data from terrestrial heavy-ion reactions}
In this section, we first outline the theoretical tools one uses
to extract information about the EOS of neutron-rich nuclear
matter from heavy-ion collisions. We put the special emphasis on
exploring the density-dependence of the symmetry energy as the
study on the EOS of symmetric nuclear matter with heavy-ion
reactions is better known to the astrophysical community and it
has been extensively reviewed, see e.g.,
Refs.~\citep{Danielewicz:2002pu,Steiner:2004fi,Lattimer:2007} for
recent reviews. We will then summarize the latest constraints on
the density dependence of the symmetry energy extracted from
studying isospin diffusion and isoscaling in heavy-ion reactions
at intermediate energies. Finally, we address the question of what
kind of isospin-asymmetry, especially for dense matter, can be
reached in heavy-ion reactions.

Heavy-ion reactions are a unique means to create in terrestrial
laboratories dense nuclear matter similar to those found in the
core of neutron stars. Depending on the beam energy, impact
parameter and the reaction system, various hadrons and/or partons
may be created during the reaction. To extract information about
the EOS of dense matter from heavy-ion reactions requires careful
modelling of the reaction dynamics and selection of sensitive
observables. Among the available tools, computer simulations based
on the Boltzmann-Uehling-Uhlenbeck ({\sc buu}) transport theory
have been very useful, see, e.g.,
Refs.~\citep{Bertsch,Danielewicz:2002pu} for reviews. The
evolution of the phase space distribution function
$f_i(\vec{r},\vec{p},t)$ of nucleon $i$ is governed by both the
mean field potential $U$ and the collision integral
$I_{collision}$ via the {\sc buu} equation
\begin{equation}
\frac {\partial f_i}{\partial t} + {\vec \nabla}_p U \cdot {\vec
\nabla}_r f_i - {\vec \nabla}_r U \cdot {\vec \nabla}_p f_i =
I_{collision}.
\end{equation}
Normally, effects of the collision integral $I_{collision}$ via
both elastic and inelastic channels including particle
productions, such as pions, are modelled via Monte Carlo sampling
using either free-space experimental data or calculated in-medium
cross sections for the elementary hadron-hadron
scatterings~\citep{Bertsch}. Information about the EOS is obtained
from the underlying mean-field potential U which is an input to
the transport model. By comparing experimental data on some
carefully selected observables with transport model predictions
using different mean-field potentials corresponding to various
EOSs, one can then constrain the corresponding EOS. The specific
constrains on the density dependence of the nuclear symmetry
energy that we are using in this work were obtained by analyzing
the isospin diffusion data~\citep{Tsang:2004} within the IBUU04
version of an isospin and momentum dependent transport
model~\citep{IBUU04}. In this model, an isospin and
momentum-dependent interaction (MDI)~\citep{Das:2002fr} is used.
With this interaction, the potential energy density $V(\rho
,T,\delta )$ at total density $\rho $, temperature $T$ and isospin
asymmetry $\delta$ is
\begin{eqnarray}
V(\rho ,T,\delta ) &=&\frac{A_{u}\rho _{n}\rho _{p}}{\rho _{0}}+\frac{A_{l}}{%
2\rho _{0}}(\rho _{n}^{2}+\rho _{p}^{2})+\frac{B}{\sigma
+1}\frac{\rho ^{\sigma +1}}{\rho _{0}^{\sigma }}(1-x\delta ^{2}) \nonumber \\
&+&\sum_{\tau ,\tau ^{\prime }}\frac{C_{\tau ,\tau ^{\prime
}}}{\rho_0}
\int \int d^{3}pd^{3}p^{\prime }\frac{f_{\tau }(\vec{r},\vec{p}%
)f_{\tau ^{\prime }}(\vec{r},\vec{p}^{\prime
})}{1+(\vec{p}-\vec{p}^{\prime })^{2}/\Lambda ^{2}}\label{MDIV}
\end{eqnarray}
In the mean field approximation, Eq. (\ref{MDIV}) leads to the
following single particle potential for a nucleon with momentum $\vec{p}$ and isospin $%
\tau $
\begin{eqnarray}
U_{\tau }(\rho ,T,\delta ,\vec{p},x)&=&A_{u}(x)\frac{\rho _{-\tau
}}{\rho_{0}}+A_{l}(x)\frac{\rho _{\tau }}{\rho _{0}}+B\left(
\frac{\rho}{\rho _{0}}\right) ^{\sigma }(1-x\delta ^{2})\nonumber\\
&-&8\tau x \frac{B}{\sigma +1}\frac{\rho ^{\sigma
-1}}{\rho_{0}^{\sigma}}\delta \rho_{-\tau }+\sum_{t=\tau ,-\tau }\frac{2C_{\tau ,t}}{\rho _{0}}\int d^{3}\vec{p}%
^{\prime
}\frac{f_{t}(\vec{r},\vec{p}^{\prime})}{1+(\vec{p}-\vec{p}^{\prime
})^{2}/\Lambda ^{2}},\label{MDIU}
\end{eqnarray}
where $\tau =1/2$ ($-1/2$) for neutrons (protons), $x$,
$A_{u}(x)$, $A_{\ell }(x)$, $B$, $C_{\tau ,\tau }$,$C_{\tau ,-\tau
}$, $\sigma $, and $\Lambda $ are all parameters given in Ref.
\citep{Das:2002fr}. The last two terms in Eq. (\ref{MDIU}) contain
the momentum dependence of the single-particle potential,
including that of the symmetry potential if one allows for
different interaction strength parameters $C_{\tau ,-\tau }$ and
$C_{\tau ,\tau }$ for a nucleon of isospin $\tau $ interacting,
respectively, with unlike and like nucleons in the background
fields. It is worth mentioning that the nucleon isoscalar
potential estimated from $U_{isoscalar}\approx (U_{n}+U_{p})/2$
agrees with the prediction of variational many-body calculations
for symmetric nuclear matter \citep{wiringa} in a broad density
and momentum range \citep{IBUU04}. Moreover, the EOS of symmetric
nuclear matter for this interaction is consistent with that
extracted from the available data on collective flow and particle
production in relativistic heavy-ion collisions up to five times
the normal nuclear matter~\citep{Danielewicz:2002pu,KLW2}. On the
other hand, the corresponding isovector (symmetry) potential can
be estimated from $U_{sym}\approx (U_{n}-U_{p})/2\delta $. At
normal nuclear matter density, the MDI symmetry potential agrees
very well with the Lane potential extracted from nucleon-nucleus
and (n,p) charge exchange reactions available for nucleon kinetic
energies up to about $100$ MeV~\citep{IBUU04}. At abnormal
densities and higher nucleon energies, however, there is no
experimental constrain on the symmetry potential available at
present.

The different $x$ values in the MDI interaction are introduced to
vary the density dependence of the nuclear symmetry energy while
keeping other properties of the nuclear equation of state fixed.
Specifically, choosing the incompressibility $K_{0}$ of cold
symmetric nuclear matter at saturation density $\rho _{0}$ to be
$211$ MeV leads to the dependence of the parameters $A_{u}$ and
$A_{l}$ on the $x$ parameter according to
\begin{eqnarray}
A_{u}(x)=-95.98-x\frac{2B}{\sigma
+1},~A_{l}(x)=-120.57+x\frac{2B}{\sigma +1},
\end{eqnarray}
with $B=106.35~{\rm MeV}$.

\begin{figure}[t!]
\centering
\includegraphics[scale=0.9]{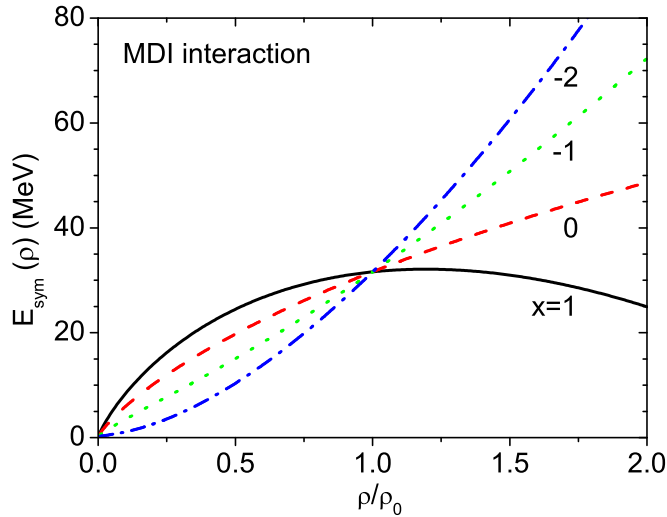}
\caption{The density dependence of the nuclear symmetry energy for
different values of the parameter $x$ in the MDI interaction.
Taken from \citep{Li05}.} \label{MDIsymE}
\end{figure}

With the potential contribution in Eq.~\ref{MDIV} and the
well-known contribution from nucleon kinetic energies in the Fermi
gas model, the EOS and the symmetry energy at zero temperature can
be easily obtained. As shown in Fig.~\ref{MDIsymE}, adjusting the
parameter $x$ leads to a broad range of the density dependence of
the nuclear symmetry energy, similar to those predicted by various
microscopic and/or phenomenological many-body theories. As
demonstrated by \citet{Li:2005jy} and \citet{Li:2005sr}, only
equations of state with $x$ between -1 and 0 have symmetry
energies in the sub-saturation density region consistent with the
isospin diffusion data and the available measurements of the skin
thickness of $^{208}Pb$ using hadronic probes. Moreover, it is
interesting to note that the symmetry energy extracted very
recently from the isoscaling analyses of heavy-ion reactions is
consistent with the MDI calculation using $x=0$
\citep{Shetty:2007}. The $E_{sym}(\rho)$ with $x=0$ is also
consistent with the RMF prediction using the FSUGold
interaction~\citep{Piek07}. We thus consider only the two limiting
cases with $x=0$ and $x=-1$ as boundaries of the symmetry energy
consistent with the available terrestrial nuclear laboratory data.

\begin{table}[t!]
\caption{{\protect\small The parameters }$F${\protect\small \ (MeV),
}$G$
{\protect\small , }$K_{sym}${\protect\small \ (MeV), } $L$%
{\protect\small \ (MeV), and }$K_{asy}${\protect\small \ (MeV) for
different values of} $x${\protect\small. Taken from
\citep{Chen:2005a}.}}
\begin{center}
\begin{tabular}{ccccccc}\label{tab.1}
$x$ & \quad $F$ & $G$ & $K_{sym}$ & $L$ & $K_{asy}$ &  \\
\hline\hline
$1$  & $107.232$ & $1.246$ & $-270.4$ & $16.4$  & -368.8 &  \\
$0$  & $129.981$ & $1.059$ & $-88.6$  & $62.1$  & -461.2 &  \\
$-1$ & $3.673$   & $1.569$ & $94.1$   & $107.4$ & -550.3 &  \\
$-2$ & $-38.395$ & $1.416$ & $276.3$  & $153.0$ & -641.7 &  \\
\hline
\end{tabular}
\end{center}
\end{table}

To ease comparisons with other models in the literature, it is
useful to parameterize the $E_{sym}(\rho )$ from the MDI interaction
and list its characteristics. Within phenomenological models it is
customary to separate the symmetry energy into the kinetic and
potential parts, see, e.g.~\citep{Pra88b},
\begin{equation}
E_{sym}(\rho )=(2^{2/3}-1)\frac{3}{5}E_{F}^{0}(\rho /\rho
_{0})^{2/3}+E_{sym}^{\mathrm{pot}}(\rho).
\end{equation}
With the MDI interaction, the potential part of the nuclear
symmetry energy can be well parameterized by
\begin{equation}
E_{sym}^{\mathrm{pot}}(\rho ) =F(x)\rho /\rho_{0} +(18.6-F(x))(\rho
/\rho _{0})^{G(x)},
\end{equation}
with $F(x)$ and $G(x)$ given in Table \ref{tab.1} for $x=1$, $0$,
$-1$ and $-2$. The MDI parameterizations for the
$E_{sym}^{\mathrm{pot}}(\rho)$ is similar but significantly
different from those used by~\citet{Pra88b}. Also shown in Table
\ref{tab.1} are other characteristics of the symmetry energy,
including its slope parameter $L$ and curvature parameter $K_{sym}$
at $\rho_0$, as well as the isospin-dependent part
$K_{\mathrm{asy}}$ of the isobaric incompressibility of asymmetric
nuclear matter~\citep{Chen:2005a}. The symmetry energy in the
subsaturation density region with x=0 and -1 can be roughly
approximately by $E_{sym}(\rho)\approx 31.6 (\rho/\rho_0)^{0.69}$
and $E_{sym}(\rho)\approx 31.6 (\rho/\rho_0)^{1.05}$, respectively.

The MDI EOS has been recently applied to constrain the mass-radius
correlations of both static and rapidly rotating neutron
stars~\citep{Li:2005sr,KLW2}. In addition, it has been also used
to constrain a possible time variation of the gravitational
constant $G$~\citep{Krastev:2007en} via the gravitochemical
heating formalism developed by \citet{Jofre:2006ug}. For
comparisons, in this work we apply also EOSs from variational
calculations with the $A18+\delta\upsilon+UIX*$ interaction (APR)
\citet{Akmal:1998cf}, and recent Dirac-Brueckner-Hartree-Fock
(DBHF) calculations~\citep{Alonso:2003aq,Krastev:2006ii}
(DBHF+Bonn B) with Bonn B One-Boson-Exchange (OBE)
potential~\citep{Machleidt:1989}. Below the baryon density of
approximately $0.07fm^{-3}$ the equations of state are
supplemented by a crustal EOS, which is more suitable for the low
density regime. Namely, we apply the EOS by~\citet{PRL1995} for
the inner crust and the one by~\citet{HP1994} for the outer crust.
At the highest densities we assume a continuous functional for the
EOSs employed in this work. (See~\citep{Krastev:2006ii} for a
detailed description of the extrapolation procedure for the
DBHF+Bonn B EOS.) The saturation properties of the nuclear
equations of state used in this paper are summarized in Table
\ref{tab.2}.
\begin{table}[t!]
\caption{Saturation properties of the nuclear EOSs (for symmetric
nuclear matter) employed in this work. Taken from \citep{KLW2}.}
\begin{center}
\begin{tabular}{lccccc}\label{tab.2}
EOS &  $\rho_0(fm^{-3})$ & $E_s(MeV)$ & $\kappa(MeV)$ & $e_{sym}(\rho_0)(MeV)$ & $m^*(\rho_0)/m$\\
\hline\hline
MDI(x=0)    & 0.160 & -16.08 & 211.00 & 31.62 & 0.67\\
MDI(x=-1)   & 0.160 & -16.08 & 211.00 & 31.62 & 0.67\\
APR         & 0.160 & -16.00 & 266.00 & 32.60 & 0.70\\
DBHF+Bonn B & 0.185 & -16.14 & 259.04 & 33.71 & 0.65\\
\hline
\end{tabular}
\end{center}
{\small The first column identifies the equation of state. The
remaining columns exhibit the following quantities at the nuclear
saturation density: saturation (baryon) density;
energy-per-particle; compression modulus; symmetry energy; nucleon
effective mass to {\it average} nucleon mass ratio (with
$m=938.926MeV$ $c^{-2}$).}
\end{table}

What is the maximum isospin asymmetry reached, especially in the
supra-normal density regions, in typical heavy-ion reactions? How
does it depend on the symmetry energy? Do both the density and
isospin asymmetry reached have to be high simultaneously in order to
probe the symmetry energy at supra-normal densities with heavy-ion
reactions? The answers to these questions are important for us to
better understand the advantages and limitations of using heavy-ion
reactions to probe the EOS of neutron-rich nuclear matter and
properly evaluate their impacts on astrophysics.

\begin{figure}[t!]
\centering
\includegraphics[scale=0.6]{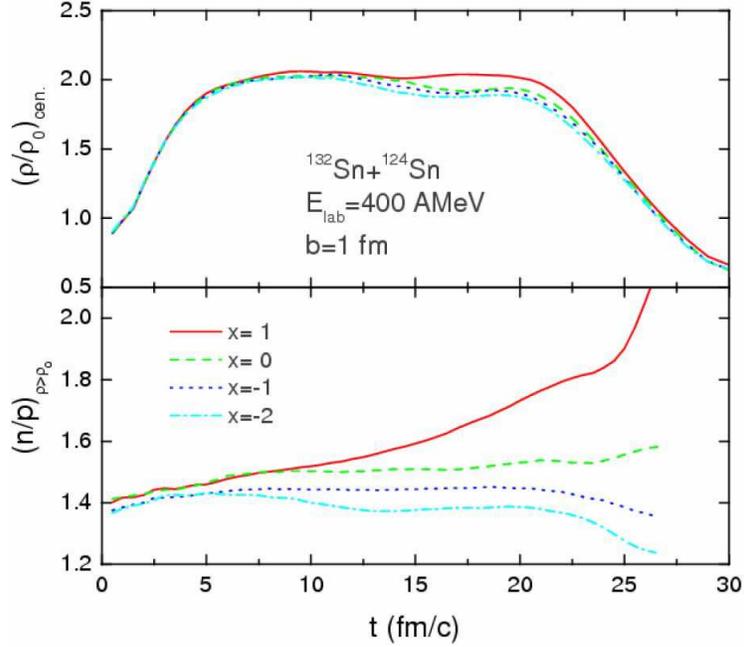}
\caption{Central baryon density (upper panel) and isospin asymmetry
(lower panel) of high density region in the reaction of $^{132}{\rm
Sn}+^{124}{\rm Sn}$ at a beam energy of 400 MeV/nucleon and an
impact parameter of 1 fm. Taken from Ref. \protect\citep{Li05}.}
\label{CentDen}
\end{figure}

To answer these questions we first show in Fig.~\ref{CentDen} the
central baryon density (upper window) and the average
$(n/p)_{\rho\geq \rho_0}$ ratio (lower window) of all regions with
baryon densities {\it higher than $\rho_0$} in the reaction of
$^{132}Sn+^{124}Sn$ at a beam energy of 400 MeV/nucleon and an
impact parameter of 1 fm. It is seen that the maximum baryon
density is about 2 times normal nuclear matter density. Moreover,
the compression is rather insensitive to the symmetry energy
because the latter is relatively small compared to the EOS of
symmetric nuclear matter around this density. The high density
phase lasts for about 15 fm/c from 5 to 20 fm/c for this reaction.
It is interesting to see in the lower window that the isospin
asymmetry of the high density region is quite sensitive to the
density dependence of the symmetry energy used in the calculation.
The soft (e.g., $x=1$) symmetry energy leads to a significantly
higher value of $(n/p)_{\rho\geq \rho_0}$ than the stiff one
(e.g., $x=-2$). This is consistent with the well-known isospin
fractionation phenomenon in asymmetric nuclear
matter~\citep{Mul95,LiKo}. Because of the $E_{sym}(\rho)\delta^2$
term in the EOS of asymmetric nuclear matter, it is energetically
more favorable to have a higher isospin asymmetry $\delta$ in the
high density region with a softer symmetry energy functional
$E_{sym}(\rho)$. In the supra-normal density region, as shown in
Fig.~\ref{MDIsymE}, the symmetry energy changes from being soft to
stiff when the parameter $x$ varies from 1 to -2. Thus the value
of $(n/p)_{\rho\ge \rho_0}$ becomes lower as the parameter $x$
changes from 1 to -2. It is worth mentioning that the initial
value of the quantity $(n/p)_{\rho\ge \rho_0}$ is about 1.4 which
is less than the average n/p ratio of 1.56 of the reaction system.
This is because of the neutron-skins of the colliding nuclei,
especially that of the projectile $^{132}Sn$. In the neutron-rich
nuclei, the n/p ratio on the low-density surface is much higher
than that in their interior. Also because of the
$E_{sym}(\rho)\delta^2$ term in the EOS, the isospin-asymmetry in
the low density region is much lower than the supra-normal density
region as long as the symmetry increases with density. In fact, as
shown in Fig. 2 of Ref.~\citep{Li:2005sr}, the isospin-asymmetry
of the low density region can become much higher than the isospin
asymmetry of the reaction system.

It is clearly seen that the dense region can become either
neutron-richer or neutron-poorer with respect to the initial state
depending on the symmetry energy functional $E_{sym}(\rho)$ used. As
long as the symmetry energy increases with the increasing density,
the isospin asymmetry of the supra-normal density region is always
lower than the isospin asymmetry of the reaction system. Thus, even
with radioactive beams, the supra-normal density region can not be
both dense and neutron-rich simultaneously, unlike the situation in
the core of neutron stars, unless the symmetry energy starts
decreasing at high densities. The high density behavior of the
symmetry energy is probably among the most uncertain properties of
dense matter as stressed by~\citep{Kut94,Kut00}. Indeed, some
predictions show that the symmetry energy can decrease with
increasing density above certain density and may even finally
becomes negative. This extreme behavior was first predicted by some
microscopic many-body theories, see e.g.,
Refs.~\citep{Pan72,Wir88a,Krastev:2006ii}. It has also been shown
that the symmetry energy can become negative at various high
densities within the Hartree-Fock approach using the original Gogny
force~\citep{Cha97}, the density-dependent M3Y interaction~
\citep{Kho96,Bas07} and about 2/3 of the 87 Skyrme interactions that
have been widely used in the literature~\citep{Sto03}. The mechanism
and physical meaning of a negative symmetry energy are still under
debate and certainly deserve more studies.

Isospin effects in heavy-ion reactions are determined mainly by the
$E_{sym}(\rho)\delta^2$ term in the EOS. One expects a larger effect
if the isospin-asymmetry is higher. Thus, ideally, one would like to
have situations where both the density and isospin asymmetry are
sufficiently high simultaneously as in the cores of neutron stars in
order to observe the strongest effects due to the symmetry energy at
supra-normal densities. However, since it is the product of the
symmetry energy and the isospin-asymmetry that matters, one can
still probe the symmetry energy at high densities where the isospin
asymmetry is generally low with symmetry energy functionals that
increase with density. Therefore, even if the high density region
may not be as neutron-rich as in neutron stars, heavy-ion collisions
can still be used to probe the symmetry energy at high densities
useful for studying properties of neutron stars.

\section{The moment of inertia of neutron stars}

Employing the EOSs described briefly in Section 2, we compute the
neutron star moment of inertia with the $RNS$\footnote{Thanks to
Nikolaos Stergioulas the $RNS$ code is available as a public domain
program at http://www.gravity.phys.uwm.edu/rns/} code developed and
made available to the public by Nikolaos
Stergioulas~\citep{Stergioulas:1994ea}. The code solves the
hydrostatic and Einstein's field equations for mass distributions
rotating rigidly under the assumption of stationary and axial
symmetry about the rotational axis, and reflectional symmetry about
the equatorial plane. $RNS$ calculates the angular momentum $J$ as
\citep{Stergioulas:2003yp}
\begin{equation}\label{eq.1}
J=\int T^{\mu\nu}\xi^{\nu}_{(\phi)}dV,
\end{equation}
where $T^{\mu\nu}$ is the energy-momentum tensor of stellar matter
\begin{equation}\label{eq.2}
T^{\mu\nu} = (\epsilon+P)u^{\mu}u^{\nu}+Pg^{\mu\nu},
\end{equation}
$\xi^{\nu}_{(\phi)}$ is the Killing vector in azimuthal direction
reflecting axial symmetry, and $dV=\sqrt{-g}d^3x$ is a proper
3-volume element ($g\equiv \det(g_{\alpha\beta})$ is the determinant
of the 3-metric). In Eq.~(\ref{eq.2}) $P$ is the pressure,
$\epsilon$ is the mass-energy density, and $u^\mu$ is the unit
time-like four-velocity satisfying $u^{\mu}u_{\mu}=-1$. For
axial-symmetric stars it takes the form $u^{\mu}=u^t(1,0,0,\Omega)$,
where $\Omega$ is the star's angular velocity. Under this condition
Eq.~(\ref{eq.1}) reduces to
\begin{equation}\label{eq.3}
J=\int (\epsilon+P)u^t(g_{\phi\phi}u^{\phi}+g_{\phi
t}u^t)\sqrt{-g}d^3x
\end{equation}
It should be noted that the moment of inertia cannot be calculated
directly as an integral quantity over the source of gravitational
field \citep{Stergioulas:2003yp}. In addition, there exists no
unique generalization of the Newtonian definition of the moment of
inertia in General Relativity and therefore $I=J/\Omega$ is a
natural choice for calculating this important quantity.

For rotational frequencies much lower than the Kepler frequency (the
highest possible rotational rate supported by a given EOS), i.e.
$\nu/\nu_k<<1$ ($\nu=\Omega/(2\pi)$), the deviations from spherical
symmetry are very small, so that the moment of inertia can be
approximated from spherical stellar models. In what follows we
review briefly this slow-rotation approximation, see e.g.
\citep{Hartle:1967he}. In the slow-rotational limit the metric can
be written in spherical coordinates as (in geometrized units
$G=c=1$)
\begin{equation}\label{eq.4}
ds^2=-e^{2\phi(r)}dt^2+\left(1-\frac{2m(r)}{r}\right)^{-1}dr^2-2\omega
r^2\sin^2\theta dt d\phi+r^2(d\theta^2+\sin^2\theta d\phi^2)
\end{equation}
In the above equation $m(r)$ is the total gravitational mass within
radius $r$ satisfying the usual equation
\begin{equation}\label{eq.5}
\frac{dm(r)}{dr}=4\pi\epsilon(r)r^{2}
\end{equation}
and $\omega(r)\equiv(d\phi/dt)_{ZAMO}$ is the Lense-Thirring angular
velocity of a zero-angular-momentum observer (ZAMO). Up to first
order in $\omega$ all metric functions remain spherically symmetric
and depend only on $r$ \citep{Morrison:2004}. In the stellar
interior the Einstein's field equations reduce to
\begin{equation}\label{eq.6}
\frac{d\phi(r)}{dr}=m(r)\left[1+\frac{4\pi
r^3P(r)}{m(r)}\right]\left[1-\frac{2m(r)}{r}\right]^{-1}\quad
(r<R_{star})
\end{equation}
and
\begin{equation}\label{eq.7}
\frac{1}{r^3}\frac{d}{dr}\left(r^4j(r)\frac{d\bar{\omega}(r)}{dr}\right)+4\frac{dj(r)}{dr}\bar{\omega}(r)=0\quad
(r<R_{star}),
\end{equation}
with $\bar{\omega}\equiv \Omega-\omega$ the dragging angular
velocity (the angular velocity of the star relative to a local
inertial frame rotating at $\omega$) and
\begin{equation}\label{eq.8}
j\equiv\left(1-\frac{2m(r)}{r}\right)^{1/2}e^{-\phi(r)}
\end{equation}
Outside the star the metric functions become
\begin{equation}\label{eq.9}
e^{2\phi}=\left(1-\frac{2M}{r}\right)\quad (r>R_{star})
\end{equation}
and
\begin{equation}\label{eq.10}
\omega=\frac{2J}{r^3}\quad (r>R_{star}),
\end{equation}
where $M=m(r=R)=4\pi\int_0^R\epsilon(r')r'^2dr'$ is the total
gravitational mass and $R$ is the stellar radius defined as the
radius at which the pressure drops to zero ($P(r=R)=0$). At the
star's surface the interior and exterior solutions are matched by
satisfying the appropriate boundary conditions
\begin{equation}\label{eq.11}
\bar{\omega}(R)=\Omega-\frac{R}{3}\left(\frac{d\bar{\omega}}{dr}\right)_{r=R}
\end{equation}
and
\begin{equation}\label{eq.12}
\phi(r)=\frac{1}{2}\ln\left(1-\frac{2M}{R}\right)
\end{equation}
The moment of inertia $I=J/\Omega$ then can be computed from
Eq.~(\ref{eq.3}). With $\Omega=u^{\phi}/u^{t}$ and retaining only
first order terms in $\omega$ and $\Omega$, the moment of inertia
reads \citep{Morrison:2004,Lattimer:2000kb}
\begin{equation}\label{eq.13}
I\approx\frac{8\pi}{3}\int^R_0(\epsilon+P)e^{-\phi(r)}\left[1-\frac{2m(r)}{r}\right]^{-1}
\frac{\bar{\omega}}{\Omega}r^4dr
\end{equation}
This slow-rotation approximation for the neutron-star moment of
inertia neglects deviations from spherical symmetry and is
independent of the angular velocity $\Omega$ \citep{Morrison:2004}.
For neutron stars with masses greater than $1M_{\sun}$
\citet{Lattimer:2005} found that, for slow-rotations, the momenta of
inertia computed through the above formalism (Eq.~(\ref{eq.13})) can
be approximated very well by the following empirical relation:
\begin{equation}\label{eq.14}
I\approx(0.237\pm
0.008)MR^2\left[1+4.2\frac{Mkm}{M_{\sun}R}+90\left(\frac{Mkm}{M_{\sun}R}\right)^4\right]
\end{equation}
The above equation is shown \citep{Lattimer:2005} to hold for a wide
class of EOSs except for ones with appreciable degree of softening,
usually indicated by achieving a maximum mass of $\sim 1.6M_{\sun}$
or less. Since none of the EOSs employed in this paper exhibit such
pronounced softening, Eq.~(\ref{eq.14}) is a good approximation for
the momenta of inertia of {\it slowly} rotating stars.

\section{Results and discussion}

\begin{figure}[h!]
\centering
\includegraphics[totalheight=2.8in]{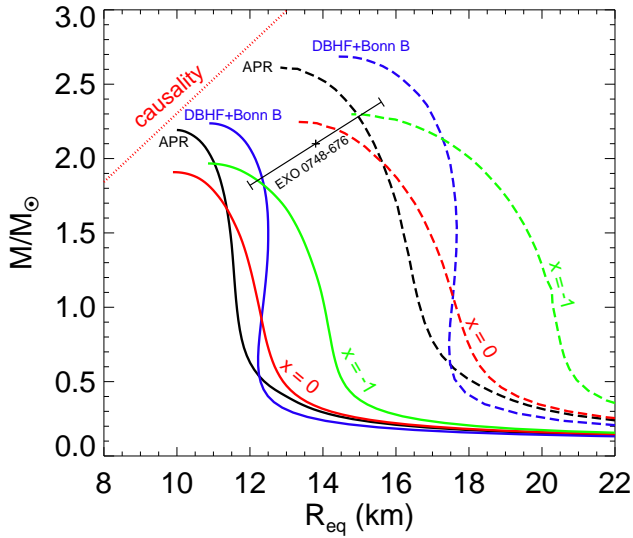}
\vspace{5mm} \caption{(Color online) Mass-radius relation for
static and maximally rotating neutron stars. Solid lines
correspond to static and broken lines to maximally rotating
stellar models. Taken form \citep{KLW2}.} \label{fig.3}
\end{figure}

We calculate the neutron star moment of inertia applying several
nucleonic EOSs (see Section 2) considering both slow- and
rapid-rotation regimes. In Fig.~\ref{fig.3} we show stellar
sequences computed with the $RNS$ code for spherical and maximally
rotating models. As seen in the figure, rapid rotation alters
significantly the mass-radius relation of rapidly rotating stars
(with respect to static configurations). Generally, for a given
EOS it increases the maximum possible mass by $\sim 15\%$, while
reducing/increasing the polar/circumferential radius by several
kilometers, leading to an overall oblate shape of the rotating
star. The degree to which the neutron star properties are impacted
by rapid rotation depends on the details of the EOS: it is greater
for models from stiffer EOS which produce less centrally condensed
and gravitational bound neutron stars \citep{1984Natur.312..255F}.
In view of these considerations, one should expect similar changes
in the moment of inertia of rapidly rotating neutron stars (with
respect to static models). We address these and other implications
next.

\subsection{Slow rotation}
\begin{figure}[b!]
\centering
\includegraphics[totalheight=3.1in]{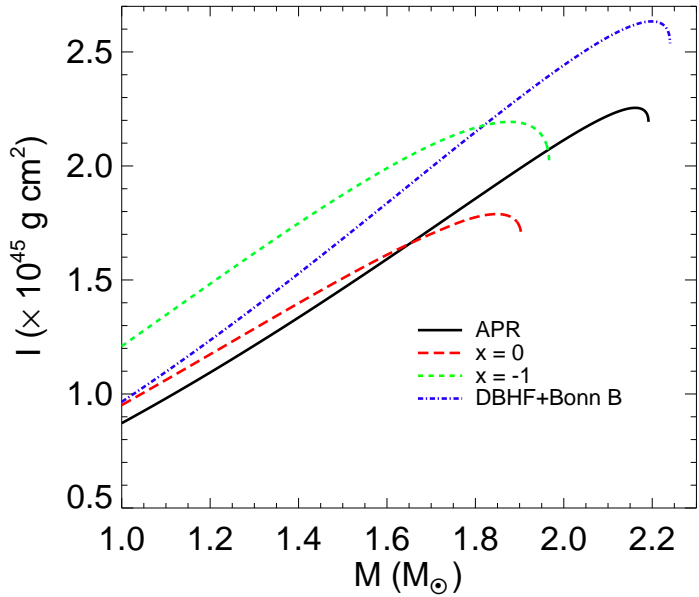}
\vspace{5mm} \caption{(Color online) Total moment of inertia of
neutron stars estimated with Eq.~(\ref{eq.14}).}
\label{fig.4}
\end{figure}
\begin{figure}[!t]
\centering
\includegraphics[totalheight=3.2in]{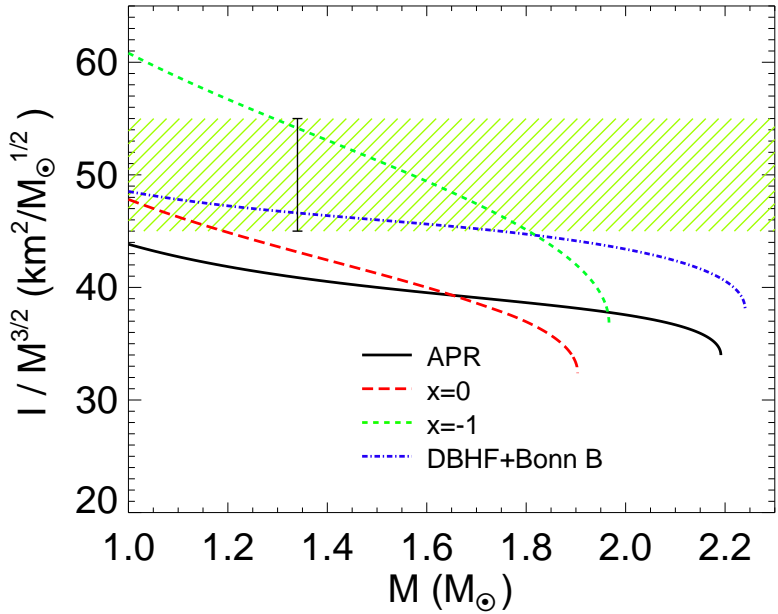}
\vspace{5mm} \caption{(Color online) The moment of inertia scaled
by $M^{3/2}$ as a function of the stellar mass $M$. The shaded
band illustrates a 10\% error of hypothetical $I/M^{3/2}$
measurement of 50 $km^2$ $M_{\sun}^{-1/2}$. The error bar shows
the specific case in which the mass is $1.34M_{\sun}$ (after
\citep{Lattimer:2005}).}
\label{fig.5}
\end{figure}

If the rotational frequency is much smaller than the Kepler
frequency, the deviations from spherical symmetry are negligible
and the moment of inertia can be calculated applying the
slow-rotation approximation discussed briefly in Section 3. For
this case \citet{Lattimer:2005} showed that the moment of inertia
can be very well approximated by Eq.~(\ref{eq.14}). In
Fig.~\ref{fig.4} we display the moment of inertia as a function of
stellar mass for slowly rotating neutron stars as computed with
the empirical relation (\ref{eq.14}). As shown in
Fig.~\ref{fig.3}, above $\sim 1.0M_{\sun}$ the neutron star radius
remains approximately constant before reaching the maximum mass
supported by a given EOS. The moment of inertia ($I\sim MR^2$)
thus increases almost linearly with stellar mass for all models.
Right before the maximum mass is achieved, the neutron star radius
starts to decrease (Fig.~\ref{fig.3}), which causes the sharp drop
in the moment of inertia observed in Fig.~\ref{fig.4}. Since $I$
is proportional to the mass and the square of the radius, it is
more sensitive to the density dependence of the nuclear symmetry
energy, which determines the neutron star radius. Here we recall
that the $x=-1$ EOS has much stiffer symmetry energy (with respect
to the one of the $x=0$ EOS), which results in neutron star models
with larger radii and, in turn, momenta of inertia. For instance,
for a ``canonical'' neutron star ($M=1.4M_{\sun}$), the difference
in the moment of inertia is more than $30\%$ with the $x=0$ and
the $x=-1$ EOSs. In Fig.~\ref{fig.5} we take another view of the
moment of inertia where $I$ is scaled by $M^{3/2}$ as a function
of the stellar mass (after \citep{Lattimer:2005}).

The discovery of the extremely relativistic binary pulsar PSR
J0737-3039A,B provides an unprecedented opportunity to test
General Relativity and physics of pulsars \citep{Burgay2003}.
\citet{Lattimer:2005} estimated that the moment of inertia of the
A component of the system should be measurable with an accuracy of
about 10\%. Given that the masses of both stars are already
accurately determined by observations, a measurement of the moment
of inertia of even one neutron star could have enormous importance
for the neutron star physics \citep{Lattimer:2005}. (The
significance of such a measurement is illustrated in
Fig.~\ref{fig.5}. As pointed by \citet{Lattimer:2005}, it is clear
that very few EOSs would survive these constraints.) Thus,
theoretical predictions of the moment of inertia are very timely.
Calculations of the moment of inertia of pulsar A
($M_A=1.338M_{\sun}$, $\nu_A=44.05Hz$) have been reported by
\citet{Morrison:2004} and \citet{BBH2005}.
\begin{table}[t!]
\caption{Numerical results for PSR J0737-3039A ($M_A=1.338M_{\sun}$,
$\nu_A=44.05Hz$).}
\begin{center}
\begin{tabular}{lcccc}\label{tab.3}
EOS &  $\epsilon_c(\times 10^{14}g$ $cm^{-3})$ & $R_{eq}(km)$ & $I(\times 10^{45}g$ $cm^2)$ & $I^{LS}(\times 10^{45}g$ $cm^2)$\\
\hline\hline
MDI(x=-1)    & 7.04 & 13.75 (13.64) & 1.63 & 1.67\\
DBHF+Bonn B  & 7.34 & 12.56 (12.47) & 1.57 & 1.43\\
MDI(x=0)     & 9.85 & 12.00 (11.90) & 1.30 & 1.34\\
APR          & 9.58 & 11.60 (11.52) & 1.25 & 1.26\\
\hline
\end{tabular}
\end{center}
{\small The first column identifies the equation of state. The
remaining columns exhibit the following quantities: central
mass-energy density, equatorial radius (the numbers in the
parenthesis are the radii of the spherical models; the deviations
from sphericity due to rotation are $\sim 1\%$), total moment of
inertia, total moment of inertia $I^{LS}$ as computed with
Eq.~(\ref{eq.14}).}
\end{table}
In Table~\ref{tab.3} we show the moment of inertia and (other
selected quantities) of PSR J0737-3039A computed with the $RNS$
code using the EOSs employed in this study. Our results with the
APR EOS are in very good agreement with those by
\citet{Morrison:2004} ($I^{APR}=1.24\times 10^{45}g$ $cm^2$) and
\citet{BBH2005} ($I^{APR}=1.23\times 10^{45}g$ $cm^2$). In the
last column of Table~\ref{tab.3} we also include results computed
with the empirical relation (Eq.~(\ref{eq.14})). From a comparison
with the results from the exact numerical calculation we conclude
that Eq.~(\ref{eq.14}) is an excellent approximation for the
moment of inertia of slowly-rotating neutron stars. (The average
uncertainty of Eq.~(\ref{eq.14}) is $\sim 2\%$, except for the
DBHF+BonnB EOS for which it is $\sim 8\%$.) Our results (with the
MDI EOS) allowed us to constrain the moment of inertia of pulsar A
to be in the range $I=(1.30-1.63)\times 10^{45}(g$ $cm^2)$.

\subsection{Rapid rotation}

\begin{figure}[!t]
\centering
\includegraphics[totalheight=2.8in]{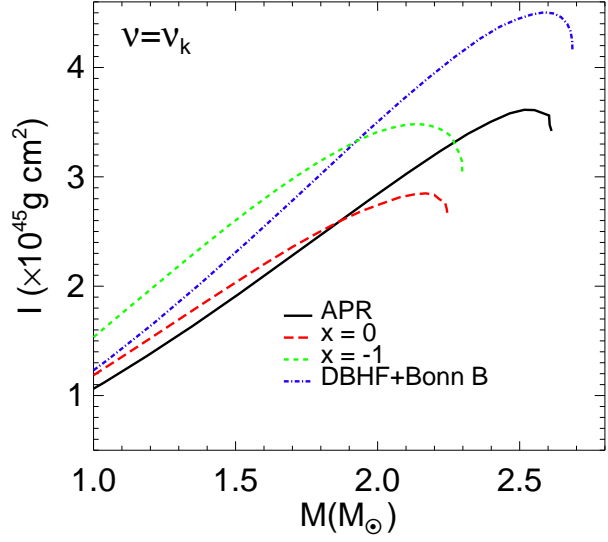}
\vspace{5mm} \caption{(Color online) Total moment of inertia for
Keplerian models. The neutron star sequences are computed with the
$RNS$ code.} \label{fig.6}
\end{figure}

\begin{figure}[!t]
\centering
\includegraphics[totalheight=3.4in]{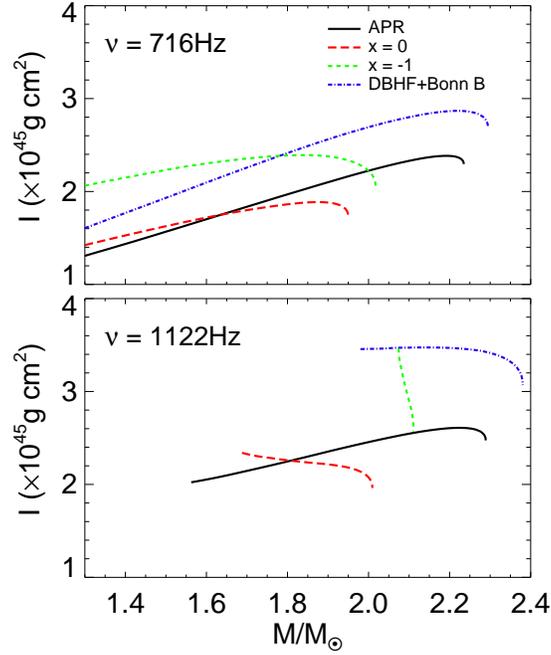}
\vspace{5mm} \caption{(Color online) Total moment of inertia as a
function of stellar mass for models rotating at 716Hz (upper
frame) and 1122 Hz (lower frame).} \label{fig.7}
\end{figure}

In this subsection we turn our attention to the moment of inertia
of rapidly rotating neutron stars. In Fig.~\ref{fig.6} we show the
moment of inertia as a function of stellar mass for neutron star
models spinning at the mass-shedding (Kepler) frequency. The
numerical calculation is performed with the $RNS$ code. We observe
that the momenta of inertia of rapidly rotating neutron stars are
significantly larger than those of slowly rotating models (for a
fixed mass). This is easily understood in terms of the increased
(equatorial) radius (Fig.~\ref{fig.3}).

\begin{figure}[!t]
\centering
\includegraphics[totalheight=2.8in]{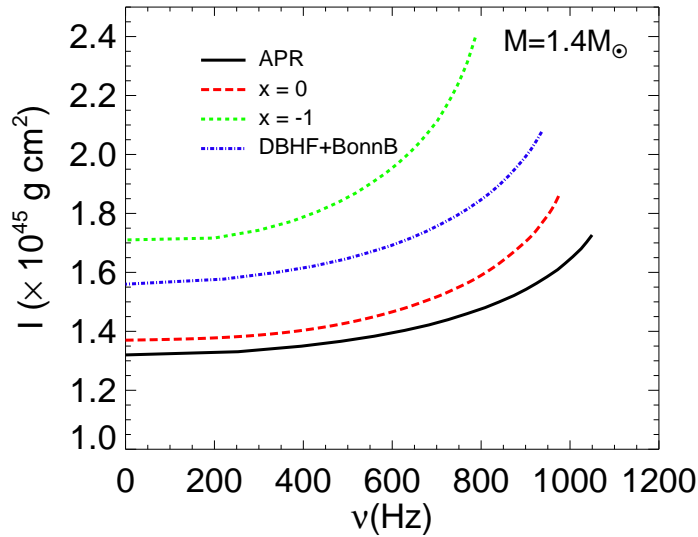}
\vspace{5mm} \caption{(Color online) Total moment of inertia as a
function of rotational frequency for stellar models with mass
$M=1.4M_{\sun}$.}\label{fig.8}
\end{figure}

We also compute the momenta of inertia of neutron stars rotating
at 716 \citep{Hessels:2006ze} and 1122Hz \citep{Kaaret:2006gr}
which are the rotational frequencies of the fastest pulsars of
today. The numerical results are presented in Fig.~\ref{fig.7}. As
demonstrated by \citet{Bejger:2006hn} and most recently by
\citet{KLW2}, the range of the allowed masses supported by a given
EOS for rapidly rotating neutron stars becomes narrower than the
one for static configurations. The effect becomes stronger with
increasing frequency and depends upon the EOS. This is also
illustrated in Fig.~\ref{fig.7}, particularly in the lower panel.
Additionally, the moment of inertia shows increase with rotational
frequency at a rate dependent upon the details of the EOS. This is
best seen in Fig.~\ref{fig.8} where we display the moment of
inertia as a function of the rotational frequency for stellar
models with a fixed mass ($M=1.4M_{\sun}$).
The neutron star sequences shown in Fig.~\ref{fig.8} are
terminated at the mass-shedding frequency. At the lowest
frequencies the moment of inertia remains roughly constant for all
EOSs (which justifies the application of the slow-rotation
approximation and Eq.~(\ref{eq.14})). As the stellar models
approach the Kepler frequency, the moment of inertia exhibits a
sharp rise . This is attributed to the large increase of the
circumferential radius as the star approaches the ``mass-shedding
point''. As pointed by \citet{1984Natur.312..255F}, properties of
rapidly rotating neutron stars display greater deviations from
those of spherically symmetric (static) stars for models computed
with stiffer EOSs. This is because such models are less centrally
condensed and gravitationally bound. This also explains why the
momenta of inertia of rapidly rotating neuron star configurations
from the $x=-1$ EOS show the greatest deviation from those of
static models.

\subsection{Fractional moment of inertia of the neutron star crust}

\begin{figure}[t!]
\centering
\includegraphics[totalheight=2.4in]{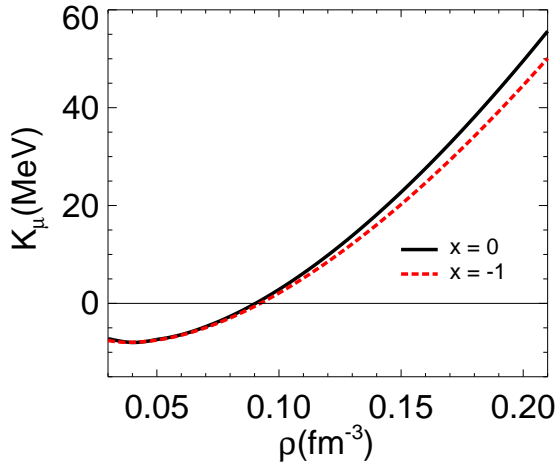}
\vspace{5mm} \caption{(Color online) The incompressibility,
$K_{\mu}$, as a function of baryon density $\rho$.}\label{fig.9}
\end{figure}

As it was discussed extensively by \citet{Lattimer:2000kb} (and
others), the neutron star crust thickness might be measurable from
observations of pulsar glitches, the occasional disrupts of the
otherwise extremely regular pulsation from magnetized, rotating
neutron stars. The canonical model of \citet{Link:1999} suggests
that glitches are due to the angular momentum transfer from
superfluid neutrons to normal matter in the neutron star crust, the
region of the star containing nuclei and nucleons that have dripped
out of nuclei. This region is bound by the neutron drip density at
which nuclei merge into uniform nucleonic matter. \citet{Link:1999}
concluded from the observations of the Vela pulsar that at least
$1.4\%$ of the total moment of inertia resides in the crust of the
Vela pulsar. For slowly rotating neutron stars, applying several
realistic hadronic EOSs that permit maximum masses of at least $\sim
1.6M_{\sun}$  \citet{Lattimer:2000kb} found that the fractional
moment of inertia, $\Delta I/I$, can be expressed approximately as
\begin{equation}\label{eq.15}
{\Delta I\over I}\simeq{28\pi P_t R^3\over
3Mc^2}{(1-1.67\beta-0.6\beta^2)\over\beta}\left[1+{2P_t(1+5\beta-14\beta^2)
\over \rho_tm_bc^2\beta^2}\right]^{-1}\
\end{equation}
In the above equation $\Delta I$ is the moment of inertia of the
neutron star crust, $I$ is the total moment of inertia, $\beta =
GM/Rc^2$ is the compactness parameter, $m_b$ is the average nucleon
mass, $\rho_t$ is the transition density at the crust-core boundary,
and $P_t$ is the transition pressure. The determination of the
transition density itself is a very complicated problem. Different
approaches often give quite different results. Similar to
determining the critical density for the spinodal decomposition for
the liquid-gas phase transition in nuclear matter, for uniform
$npe$-matter, \citet{Lattimer:2000kb} and more recently
\citet{Kubis:2007} have evaluated the crust transition density by
investigating when the incompressibility of $npe$-matter becomes
negative, i.e
\begin{equation}\label{eq.16}
K_{\mu}=\rho^2{d^2E_0\over d\rho^2}+2\rho{dE_0\over
d\rho}+\delta^2\left[\rho^2{d^2E_{sym}\over
d\rho^2}+2\rho{dE_{sym}\over
d\rho}-2E_{sym}^{-1}\left(\rho{dE_{sym}\over
d\rho}\right)^2\right]<0
\end{equation}
(see Fig.~\ref{fig.9}) where $E_0(\rho)$ is the EOS of symmetric
nuclear matter, $E_{sym}$ is the nuclear symmetry energy, and
$\delta=(\rho_n-\rho_p)/(\rho_n+\rho_p)$ is the asymmetry
parameter. Using this approach and the MDI interaction,
\citet{Kubis:2007} found the transition density of $0.119, 0.092,
0.095$ and $0.160 fm^{-3}$ for the $x$ parameter of $1, 0, -1 $
and $-2$, respectively. Similarly, we have calculated the
transition densities and pressures for the EOSs employed in this
work. Our results are summarized in Table~\ref{tab.4}. We find
good agreement between our results and those by \citet{Kubis:2007}
with the MDI interaction. It is interesting to notice that the
transition densities predicted by all EOSs are in the same density
range explored by heavy-ion reactions at intermediate energies.
The MDI interaction with $x=0$ and $x=-1$ constrained by the
available data on isospin diffusion in heavy-ion reaction at
intermediate energies thus limits the transition density rather
tightly in the range of $\rho_t=[0.091-0.093](fm^{-3})$.

\begin{table}[b!]
\caption{Transition densities and pressures for the EOSs used in
this paper.}
\begin{center}
\begin{tabular}{lcccc}\label{tab.4}
EOS &  MDI(x=0) & MDI(x=-1) & APR & DBHF+Bonn B\\
\hline\hline
$\rho_t(fm^{-3})$     & 0.091 (0.095) & 0.093 (0.092) & 0.087 & 0.100\\
$P_t(MeV$ $fm^{-3})$  & 0.645 & 0.982 & 0.513 & 0.393\\
\hline
\end{tabular}
\end{center}
{\small The first row identifies the equation of state. The
remaining rows exhibit the following quantities: transition density,
transition pressure. The numbers in the parenthesis are the
transition densities calculated by \citet{Kubis:2007}.}
\end{table}

\begin{figure}[t!]
\centering
\includegraphics[totalheight=2.5in]{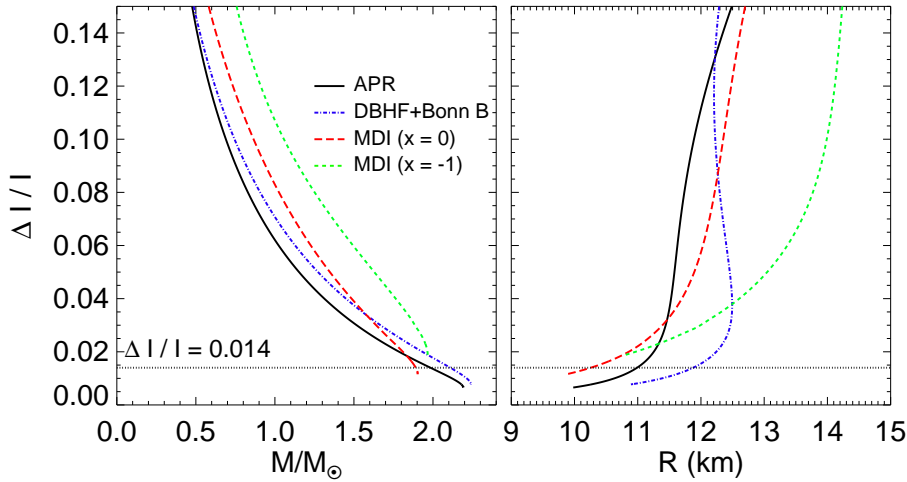}
\vspace{5mm} \caption{(Color online) The fractional moment of
inertia of the neutron star crust as a function of the neutron
star mass (left panel) and radius (right panel) estimated with
Eq.~(\ref{eq.15}). The constraint from the glitches of the Vela
pulsar is also shown.}\label{fig.10}
\end{figure}

The fractional momenta of inertia $\Delta I/I$ of the neutron star
crusts are shown in Fig.~\ref{fig.10} as computed through
Eq.~(\ref{eq.15}) with the parameters listed in Table~\ref{tab.4}.
It is seen that the condition $\Delta I/I>0.014$ extracted from
studying the glitches of the Vela pulsar does put a strict lower
limit on the radius for a given EOS. It also limits the maximum
mass to be less than about $2M_{\sun}$ for all of the EOSs
considered. Similar to the total momenta of inertia the ratio
$\Delta I/I$ changes more sensitively with the radius as the EOS
is varied.

\section{Summary}

Recent experiments in terrestrial nuclear laboratories have
narrowed down significantly the range of the EOS of neutron-rich
nuclear matter although there are still many remaining challenges
and uncertainties. In particular, the EOS for symmetric nuclear
matter was constrained up to about five times the normal nuclear
matter density by collective flow and particle production in
relativistic heavy-ion reactions. The density dependence of the
symmetry energy was constrained at subsaturation densities by
isospin diffusion and isoscaling in heavy-ion reactions at
intermediate energies. Applying the EOSs constrained by the
heavy-ion reaction data we have studied the neutron star momenta
of inertia of both slowly and rapidly rotating models within well
established formalisms. We found that the moment of inertia of PSR
J0737-3039A is limited in the range of $I=(1.30-1.63)\times
10^{45}(g$ $cm^2)$. The neutron star crust-core transition density
falls in a very narrow interval of
$\rho_t=[0.091-0.093](fm^{-3})$.  The corresponding fractional
momenta of inertia $\Delta I/I$ of the neutron star crust are also
constrained.  It is also found that the moment of inertia
increases with rotational frequency at a rate strongly dependent
upon the EOS used.

\section*{Acknowledgements}We would like to thank Lie-Wen Chen, Wei-Zhou Jiang and Jun Xu
for helpful discussions. Plamen Krastev acknowledges the
hospitality of the Institute for Nuclear Theory (INT) at the
University of Washington where parts of this work were
accomplished. This work was supported by the National Science
Foundation under Grant No. PHY0652548 and the Research Corporation
under Award No. 7123.

\end{document}